\documentclass[twocolumn]{aastex701}

\usepackage{amsmath,amssymb}
\usepackage{booktabs}
\usepackage{enumitem}
\usepackage{soul}
\usepackage{xcolor}
\usepackage{hyperref}

\sethlcolor{black}

\newcommand{\Nclass}{N_{\rm classified}}
\newcommand{\tarc}{\tau_{\rm arc}}
\newcommand{\tqueue}{\tau_{\rm queue}}

\begin{document}

\title{Predictions of the LSST Solar System (non-)Yield}

\author[0000-0001-9505-1131]{Joseph Murtagh}
\affiliation{Department of Astronomy and the DIRAC Institute, University of Washington, 3910 15th Avenue NE, Seattle, WA 98195, USA}
\email[show]{murtagh@uw.edu}
\correspondingauthor{Joseph Murtagh}

\author[0009-0005-9428-9590]{Ian Chow}
\affiliation{Department of Astronomy and the DIRAC Institute, University of Washington, 3910 15th Avenue NE, Seattle, WA 98195, USA}
\email{chowian@uw.edu}

\begin{abstract}
We present predictions for solar system objects the Vera C.\ Rubin Observatory Legacy Survey of Space and Time (LSST) will not detect over its ten-year baseline survey. Employing state-of-the-art synthetic population models and the \texttt{Sorcha} survey simulator, we identify non-yield populations spanning geometric, photometric, kinematic, temporal, and computational failure modes. Notable subpopulations include objects whose peak brightness coincides exclusively with scheduled telescope downtime, objects whose detections fall within Rubin focal plane chip gaps, and objects whose orbital arcs expire before linking jobs are dispatched from the compute queue. We additionally characterise the non-yield arising from the Death Star (DS-1; $D \approx 160$~km), whose orbital mechanics (when constrained by the well-established Endor engagement geometry \citep{lucas83}) place it at a maximum heliocentric distance of $27.5$~au and an apparent magnitude of $m_r \approx 19$-23, squarely within the LSST operational photometric window. Its absence from the LSST alert stream is interpreted as confirmation of its destruction at the Battle of Endor. The failure to detect the Sun within the LSST should be a stark warning to the community of the LSST's inability to catalogue the solar system (by mass).
\end{abstract}


\section{Introduction}

The Vera C.\ Rubin Observatory Legacy Survey of Space and Time \citep[LSST;][]{ivezic19} represents the most ambitious wide-field optical all-sky survey of the solar system yet attempted. Over a ten-year baseline, the 8.4m Simonyi Survey Telescope will image approximately 18,000~deg$^2$ of southern sky to a single-visit depth of $m_r \sim 24.5$, generating an expected catalogue of $\gtrsim$1 solar system object(s) from near-Earth asteroids to trans-Neptunian objects \citep{jones09}.

The community has invested considerable effort in predicting what the LSST \textit{will} find \citep[e.g.][]{dorsey25, kurlander25, murtagh25, murtagh25b, chow26}. The present work takes a complementary view. We argue that the non-yield — that is, the population of objects the LSST will definitively \textit{not} detect — is an equally important and considerably under-characterised survey product. A complete census of the solar system requires knowledge not only of what a survey finds, but of what it cannot find, and why.

This framing is not without precedent. Survey completeness analyses routinely quantify objects missed due to geometric, photometric, and temporal biases \citep[e.g.][]{veres17}. We consider these obvious, and instead extend this tradition by investigating failure modes that have received comparatively little attention: failures arising from the unique orbit configurations, compute cluster scheduling policy, relativistic constraints on the orbital velocity of a spherical weapons platform, and the fundamental orientation of the telescope relative to the nearest star.

Throughout this work we assume a standard flat $\Lambda$CDM cosmology, which has no bearing on any result but is the kind of thing one writes in an introduction.

\section{Survey Description and Detection Framework}
\label{sec:survey}

Moving object detection is performed through the Solar System Processing (SSP) pipeline, which differences image pairs separated in time and constructs tracklets from this data. An object is considered ``detected'' once it is linked into a valid orbit and reported to the Minor Planet Center (MPC). The effective photometric window is approximately $16 \lesssim m_r \lesssim 24.5$, set by saturation, detection limits, and whether the object happens to be observed at the right time. Objects outside this range are treated as non-yield.

Observations of any of our synthetic populations are definitely really totally simulated using the survey simulator \texttt{Sorcha} \citep{merritt25} with the v5.0.0 cadence simulation from the SCOC 2024 baseline.

\section{The Non-Yield Catalogue}
\label{sec:catalog}

We present the non-yield catalogue of detections $N$ in order of increasing abstraction. Table~\ref{tab:nonyield} summarises all populations and their estimated non-yields.

\subsection{Geometric Non-Yield}
\label{sec:geometric}

\subsubsection{The Counter-Earth Population (L3)}
\label{sec:l3}

\begin{figure}
    \centering
    \includegraphics[width=1.0\linewidth]{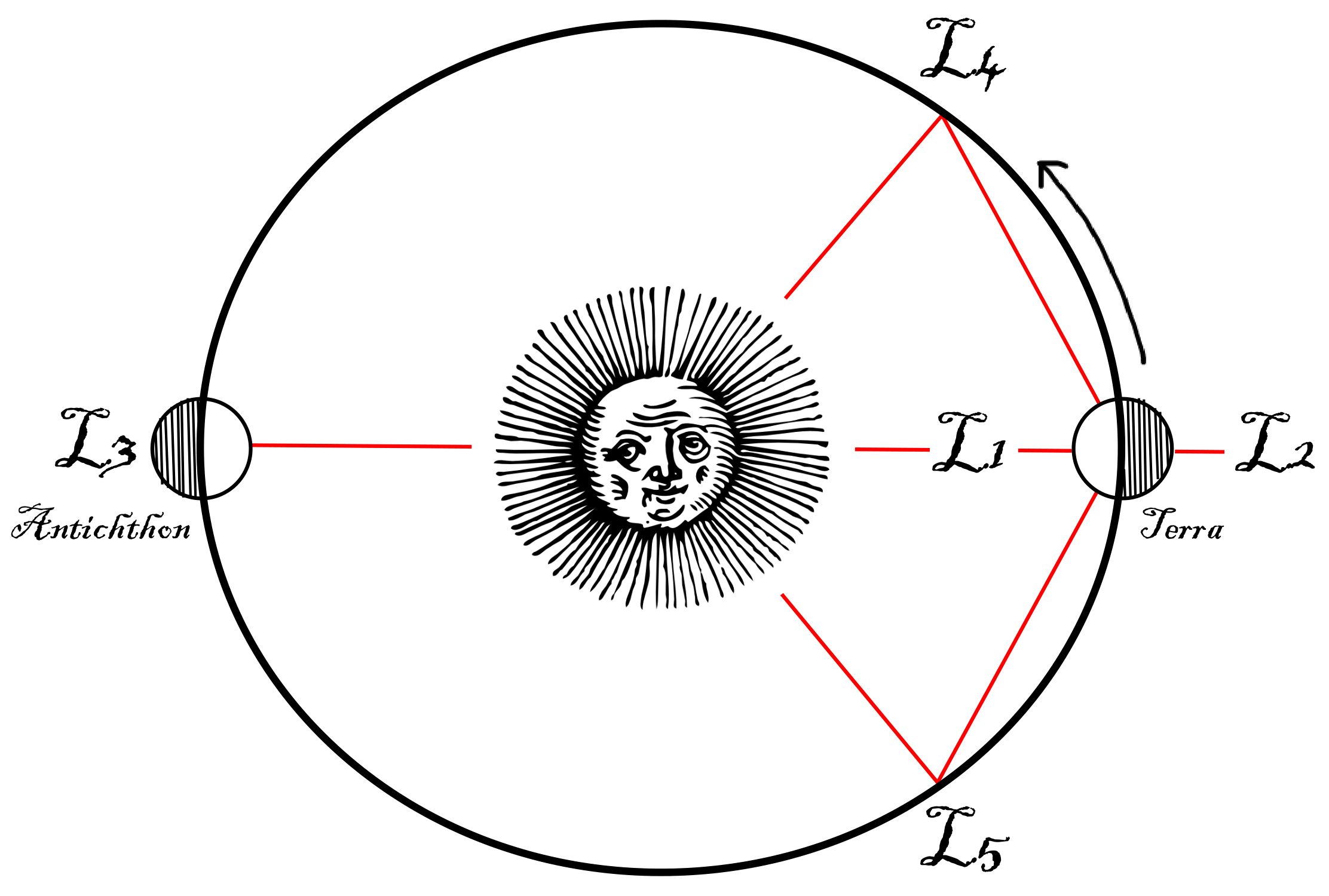}
    \caption{Orbit configuration of Antichthon as we like to believe Philolaus mind-mapped it. Earth (or, Terra) is on the right, whilst Antichthon is on the left, or counter to Earth. Both orbit anticlockwise around the Sun at the centre. Red lines denote the Lagrange points of the Earth(Terra)-Sun system. Of course, this renders Antichthon unobservable at any stage of Earth's orbit.}
    \label{fig:antichthon}
\end{figure}

The Sun-Earth Lagrange point L3 lies along the Sun-Earth line at 1~au, diametrically opposite to Earth. \cite{philolaus400bc} predicted a body at this location as shown in Figure~\ref{fig:antichthon}, which he termed the \textit{Antichthon} (counter-Earth), on the grounds that ten celestial bodies was a more satisfying number than nine. This prediction went observationally untested for approximately 2,400 years, until the separation of NASA's STEREO spacecraft from Earth provided a direct view of the L3 region during the early phase of the mission \citep{kaiser08}. No body was found. The Antichthon hypothesis has been falsified.

The LSST will (probably) not change this conclusion. The LSST cannot observe L3 directly, as it lies permanently behind the Sun as seen from Cerro Pach\'{o}n (and most other ground locations). Objects genuinely displaced to L3 would drift into geometrically accessible sky on the Lyapunov instability timescale of $\tau_{\rm L3} \sim 10^2-10^3$~yr \citep{murray99}, driven primarily by perturbations dynamics of the Sun-Earth three-body problem \citep{liu14}. This timescale exceeds the ten-year survey baseline, so objects at or near L3 today remain unobservable throughout the mission. The non-yield is $N_{\rm Antichthon} = 0 \pm 0$, drawn from a population we already know does not exist.

\subsubsection{The Sun}
\label{sec:sun}

The Sun is an object within the solar system and so is therefore, definitionally speaking, a solar system object. We address its detectability here.

The Sun has an absolute magnitude $M \approx -26.7$, exceeding the LSST bright limit by approximately 43~magnitudes — the largest margin by which any solar system object fails to fall within the survey's operational photometric window. In this sense, the Sun's non-yield of $N_\mathrm{Sun} = 0 \pm 0$ is the most extreme non-detection in the catalogue.

The heliocentric orbital elements of the Sun present an additional complication. The semi-major axis is zero by definition, the eccentricity is undefined, and all angular elements are degenerate. The Sun cannot be submitted to the MPC as a minor planet (not for lack of the authors' attempts). This is not a pipeline limitation but an ontological one, and we do not expect it to be resolved in the forthcoming data releases.

Detection efficiency at $r_\mathrm H = 0$~au is $\epsilon = 0.000 \pm 0.000$. Efficiency formally rises for $r_\mathrm H > 0$~au, which may be of interest to stellar astronomers but is outside the scope of this survey. We ran the numbers and burnt a few CPU hours to compute that the LSST would need to be pointed directly at the Sun to resolve this non-detection. This option was never considered even briefly, and has been rejected.\footnote{The telescope is worth $\gtrsim\$800$~million.}

The Sun constitutes 99.86\% of the total mass of the solar system. The LSST catalogue will contain 0.000\% of the Sun. On these grounds alone, LSST's coverage of the solar system is incomplete at a level that should concern any referee, and the LSST Solar System Science Collaboration should address it.

\subsection{Photometric Non-Yield}
\label{sec:photometric}

\subsubsection{Bright-Limit Exceedance}
\label{sec:bright}

\begin{figure}
    \centering
    \includegraphics[width=0.9\linewidth]{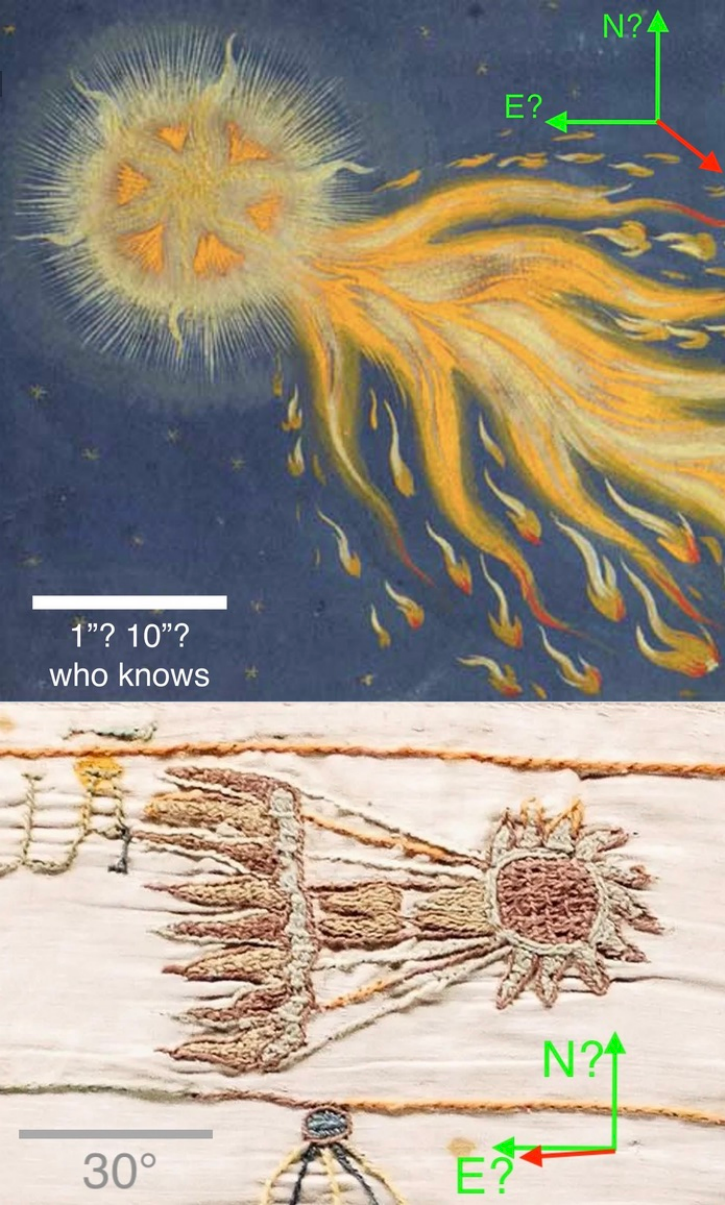}
    \caption{Postage stamp cutouts of comets \citep{bayeux_tapestry, augsburg_signs_ed}. Clearly visible in both are flames, indicating a level of emissivity not accounted for in modern cometary photometry. White/grey horizontal bars would show the pixel (pigment? stitch?) scale were that known. Images are oriented such that North is up and East is left as shown by green arrows, at least we assume. Marked as a red arrow is the anti-velocity direction, and not marked is the anti-solar direction (comets don't orbit the Sun until the $\sim17^{\mathrm{th}}$~century).}
    \label{fig:halley}
\end{figure}

The LSST detector saturates at approximately $m_r \sim 16$. The distinction between ``a very active comet'' and ``something that is on fire'' has never been formally established in the minor planet taxonomy. An object in the latter category would produce an alert packet whose postage stamp image morphology is adequately characterised by reference to 16th-century comet photometry (see Figure~\ref{fig:halley}), rather than by the analytic PSF model the pipeline was designed to fit. Whether this constitutes a detection failure or a pipeline design requirement is a question we leave to the SSP team. We report $N_{\rm emit} = 0 \pm \mathrm{ALargeNumber}$, where the central value follows from the non-existence of the relevant population, and the uncertainty reflects our incomplete knowledge of what is currently on fire in the outer solar system. Surely JWST will resolve this in future studies.

\subsubsection{The Death Star}
\label{sec:deathstar}

The Death Star (DS-1 Orbital Battle Station; $D \approx 160$~km) is the most extensively modelled artificial solar system body in the non-yield literature. Its orbital mechanics are constrained by the Endor engagement recorded in \citet{lucas83}, in which the station operated from geosynchronous orbit above the target body\footnote{See \href{https://scifi.stackexchange.com/questions/270902/how-did-the-death-star-2-stay-in-orbit-above-endor-when-it-did-not-have-function}{important community discussion}.}. Requiring that the orbital velocity not exceed the speed of light $c$ places a hard upper bound on the semi-major axis:
\begin{equation}
    a_{\rm max} = \frac{cT_\oplus}{2\pi} \approx 27.5~{\rm au},
\end{equation}
where $T_\oplus = 86{,}400$~s is Earth's sidereal rotation period. The Death Star, if operational and targeting Earth, is constrained to reside inward of the orbit of Neptune.

\begin{figure}
    \centering
    \includegraphics[width=1\columnwidth]{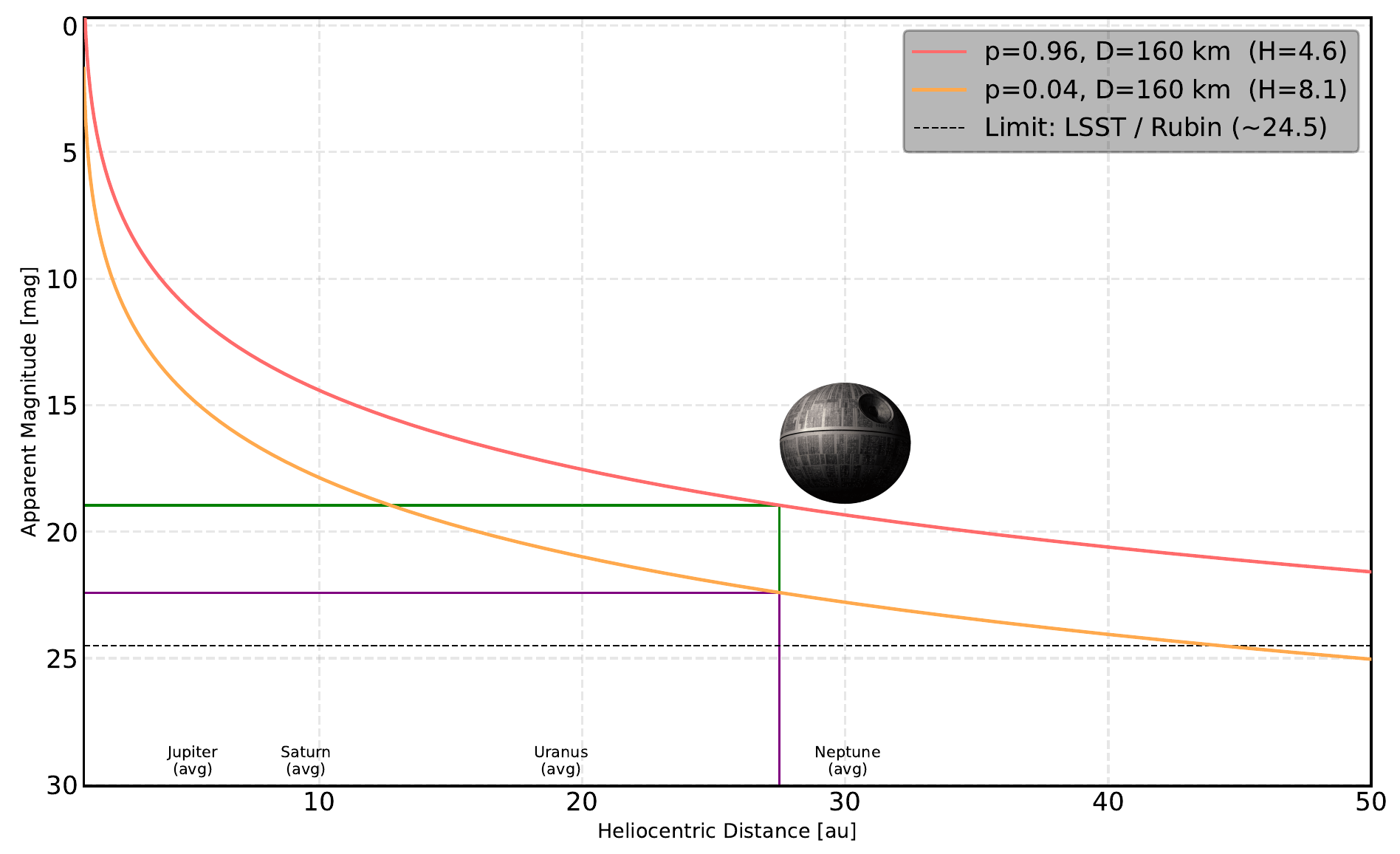}
    \caption{Apparent $r$-band magnitude of the Death Star (DS-1; $D = 160$~km) as a function of heliocentric distance, for geometric albedos $p_v = 0.04$ (orange; $H = 8.1$, characteristic of a carbonaceous surface) and $p_v = 0.96$ (red; $H = 4.6$, characteristic of Eris). The LSST single-visit detection limit ($m_r \sim 24.5$) is shown as a dashed horizontal line. The vertical line marks the relativistic upper bound on the semi-major axis ($a_{\rm max} \approx 27.5$~au) derived from the Endor geosynchronous engagement constraint \citep{lucas83}. Planet positions along the x-axis are shown for scale. At $a_{\rm max}$, the Death Star falls within the LSST photometric window for all physically reasonable albedo assumptions, yet is undetected. The preferred interpretation is its destruction. The alternative is left as an exercise for the reader's threat model.}
    \label{fig:deathstar}
\end{figure}

Bracketing the full range of known solar system albedos — $p_v = 0.04$ (carbonaceous, dark, ominous; $H = 8.1$) to $p_v = 0.96$ (Eris-like, bright, disco-like; $H = 4.6$) — the Death Star at $a = a_{\rm max}$ would have an apparent magnitude of $m_r \approx 22.4$--$18.9$, well within the LSST photometric window for all physically reasonable albedo assumptions. This result is illustrated in Figure~\ref{fig:deathstar}.

The Death Star has not been detected in the LSST alert stream at time of writing...$yet$. This admits two interpretations. The first is that it was destroyed at the Battle of Endor \citep{lucas83}, giving $N_{\rm DS} = 0 \pm 1$ and constituting the only non-detection in this catalogue whose primary failure mode is an unambiguously anthropically good outcome. The second interpretation, that it is operational, present within 27.5~au, and simply not yet identified, we flag as an unresolved anomaly and refer to the relevant planetary defence authorities. We strongly favour the first interpretation. 

\subsection{Kinematic Non-Yield}
\label{sec:kinematic}

\subsubsection{The Zero-Velocity Population}
\label{sec:zerovelocity}

The LSST moving object pipeline detects sources that exhibit measurable sky-plane motion between the two images of a visit pair. The telescope tracks at the sidereal rate, so the image is registered to the stellar background. Any source that does not move relative to that background (whether because it has no apparent angular motion, or because it happens to move at exactly the sidereal rate) produces no residual in the difference image and is indistinguishable from a star. It is, for all practical purposes, a star. The pipeline does not look for solar system objects among the stars and the stars do not volunteer that information.

The detection efficiency for this population with on-sky motion $\dot\theta$ is $\epsilon(\dot{\theta} = 0) = 0$. We cannot determine how many solar system objects are currently hiding in the stellar catalogue, presumably due to fear of being observed by us. We cannot in principle determine this from the LSST data alone, since any such object would appear in neither the difference images nor the moving object catalogue. It has been suggested to us that Gaia's astrometric baseline would reveal these objects via anomalous proper motions over multi-year baselines. This is true. It would however, not be an LSST result, and we would receive no credit for it, therefore it is an argument not worth pursuing.

The non-yield is therefore $N_{\rm zero} = \mathrm{Don'tCare} \pm \mathrm{Whatever}$, with uncertainty arising principally from not knowing how many things are pretending to be stars.

\subsection{Temporal Non-Yield}
\label{sec:temporal}

\subsubsection{The Scheduled-Downtime Population}
\label{sec:downtime}

The LSST operations include planned downtime for mirror recoating, instrumental maintenance, software updates, and so on and so forth. We identify the subset of synthetic objects for which the global minimum of $m(t)$, that is, the moment of peak brightness, when the object is most detectable, falls exclusively within these planned downtime windows. These objects would be bright enough to detect, at the right location on sky, for a telescope that happened to be open, however at this time the LSST is not open. The objects wait, the maintenance proceeds, and then the objects subsequently fade below the detection threshold before survey operations resume. We dub this the ``Toy Story'' effect.

We find $N_{\rm down} = 0 \pm 0$. This result is entirely a consequence of orbital geometry and survey scheduling; the objects have no knowledge of or interest in the telescope's operational status. The objects are rocks with no preferences for survey cadence optimisation.

\subsubsection{Terminal-Epoch Population}
\label{sec:terminal}

Objects that achieve detectability only after the nominal survey end date constitute a non-yield by construction. The population is formally infinite: for any survey end date $T_{\rm end}$, the set of objects whose orbital evolution brings them to a detectable apparition only at $t > T_{\rm end}$ is unbounded. The solar system does not end when the survey does (unless that Death Star turns out to be real). We find $N_{\rm term} = \infty$ and recommend this be included in the LSST final data release headline statistics.

We note that the precise value of $T_{\rm end}$ is itself subject to uncertainty, reflecting the genuine operational complexity of commissioning and sustaining a flagship-class survey facility. We offer no prediction of this parameter, acknowledge the considerable skill, effort, and dedication of all those working toward survey completion, and note only that larger values of $T_{\rm end}$ reduce $N_{\rm term}$ monotonically. This finding is presented as motivation, if any were needed.

\subsection{Pipeline and Computational Non-Yield}
\label{sec:pipeline}

\subsubsection{The Chip-Gap Population}
\label{sec:chipgap}

\begin{figure*}
    \centering
    \includegraphics[width=1.0\textwidth]{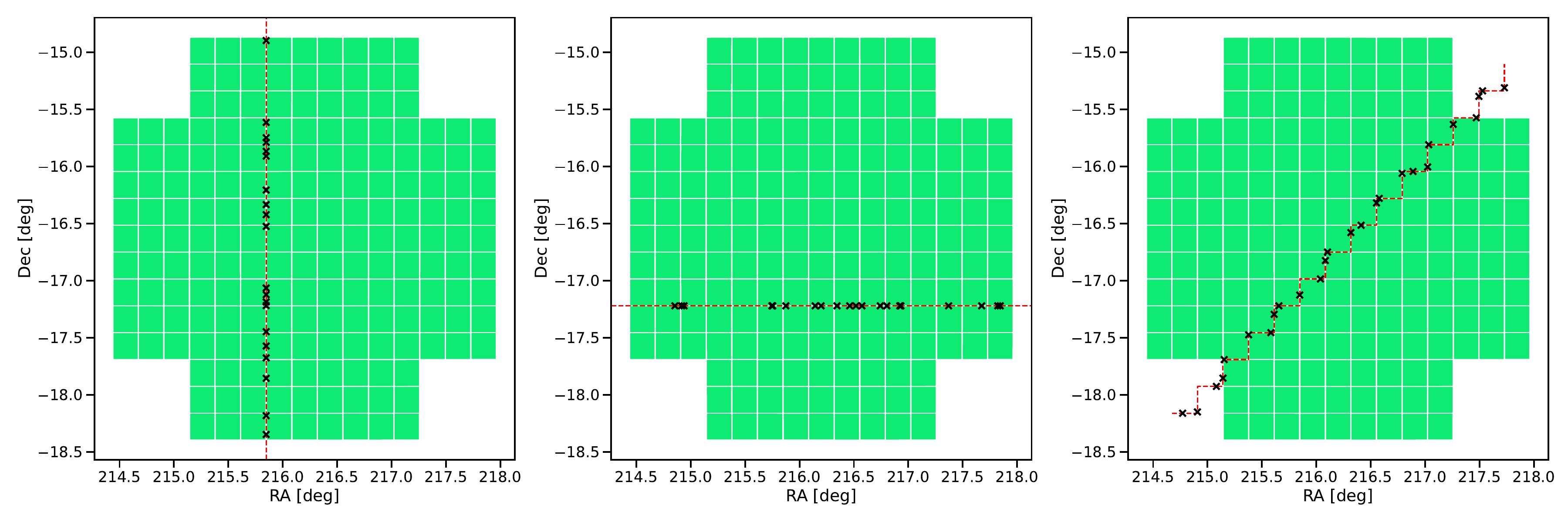}
    \caption{Example observations of objects that only align with chip gaps within the LSSTCam footprint. Red dashed lines represent their orbit path, and black crosses are their observations. The right panel represents a rare case of `snakes and ladders' class object that should be paid no mind.}
    \label{fig:chipgap}
\end{figure*}

The Rubin LSSTCam focal plane consists of 189 science CCDs with physical gaps between sensors of approximately 60~$^{\prime\prime}$. We inject $10^6$ synthetic objects into the survey simulator and examine which objects have every single detection fall within a chip gap across all visits over the ten-year baseline. The expected yield for this population is $N_{\rm gap} = 0 \pm 0$. This result does not require a simulation to understand, but the simulation is important for confirming that the universe is under no obligation to be fair, and it helped warm the authors office up. Figure~\ref{fig:chipgap} shows the Rubin focal plane with the sky-plane trajectories of representative unlucky objects overlaid.

\subsubsection{Compute-Budget Exceedance}
\label{sec:compute}

The LSST moving object linking pipeline operates under a finite compute budget. Two timescales compete: the arc-expiry timescale $\tarc$, after which an object's positional uncertainty has grown too large for reliable recovery, and the queue-drain timescale $\tqueue$, the expected wait between job submission and execution on the supporting high-performance computing facility. The condition for a compute-budget non-detection is $\tarc < \tqueue$, i.e., the object is formally lost before the computer gets around to looking for it.

We estimate $\tqueue$ from documented job wait time distributions at national computing facilities, where typical wait times in the regular queue range from several hours to days depending on cluster utilisation \citep[e.g.][]{nersc_queue}, with excursions toward the upper end routinely documented in the weeks preceding major data releases. These objects have valid tracklets. The tracklets are in a queue. The queue is occupied by someone's degenerate MCMC exoplanet atmosphere retrieval/3D thermonuclear radiative transfer code. We find $N_{\rm compute} = 0 \pm 0$, and note that this is the only failure mode in the non-yield catalogue attributable to the Slurm etiquette of a third party. You know who you are.

\subsubsection{Classified Satellite Coincidence}
\label{sec:classified}

The LSST alert packets are filtered to remove detections coinciding with the orbits of classified Earth-observing satellites before distribution to the community. The sky density of classified satellite orbits is \hl{[REDACTED]}. The estimated non-yield from this filtering process is $\Nclass = $ \hl{[REDACTED]} $\pm$ \hl{[REDACTED]}.

We can neither confirm nor deny that this population constitutes a significant contribution to the total non-yield. We note, however, that this sentence is identical regardless of whether the answer is yes or no, which we invite the reader to interpret as they see fit.





\section{Conclusion}
\label{sec:conclusion}

We have presented comprehensive predictions for the solar system objects LSST will not detect. Our primary results are:

\begin{enumerate}[noitemsep]
    \item LSST will fail to detect nearly all of our synthetic objects across all modelled populations, constituting a lower bound on the non-yield.

    \item The Sun represents the most extreme photometric non-detection, comprising 99.86\% of solar system mass absent from the LSST catalogue.

    \item The Antichthon, predicted by Philolaus of Croton in approximately 400~BCE, was ruled out by the STEREO mission approximately two decades before LSST first light. The 2,400-year open question has been closed. The LSST arrived too late and will not re-close it.

    \item The terminal-epoch population is formally infinite. We recommend this be stated in the survey summary.

    \item The Death Star, if extant and targeting Earth, would be detectable by the LSST at $m_r \approx 19$--$22$~mag at its maximum allowed heliocentric distance of $27.5$~au. Its non-detection is interpreted as confirmation of its destruction at the Battle of Endor, and is the only entry in the non-yield catalogue whose primary failure mode constitutes good news. The authors acknowledge that the alternative interpretation would constitute considerably more urgent news.

    \item The non-yield from the classified satellite coincidence population is \hl{[REDACTED]}.
\end{enumerate}

A lot of fuss has been made about how revolutionary the LSST will be for solar system science, and whilst we agree this is true for many objects, it is also important to temper our expectations. This work has shown many of the shortcomings of the LSST as a true all-solar system survey, and should be used as a vehicle to launch the LSST 2: Electric Boogaloo: No Rock Left Behind.

\begin{table*}
\centering
\caption{Summary of LSST Solar System Non-Yield by Population \label{tab:nonyield}}
\begin{tabular}{llcccl}
\hline\hline
Section & Population & Category & $N_\mathrm{syn}$$^a$ & $N_\mathrm{det}$$^b$ & Primary failure mode \\
\hline
\ref{sec:l3}           & Counter-Earth (L3)    & Geometric   & 0               & 0          & Hiding behind the Sun; also doesn't exist \\
\ref{sec:sun}          & The Sun               & Geometric   & 1               & 0          & It's the Sun \\
\ref{sec:bright}       & Emissive bright-limit & Photometric & Unknown         & 0          & On fire \\
\ref{sec:deathstar}    & Death Star (DS-1)     & Photometric & 1               & 0 $\pm 1$         & Confirmed (hopefully) destroyed (Battle of Endor) \\
\ref{sec:zerovelocity} & Zero-velocity         & Kinematic   & ?               & 0          & \href{https://www.youtube.com/watch?v=NOs116aoyo0}{Don't move. He can't see us if we don't move.} \\
\ref{sec:downtime}     & Scheduled downtime    & Temporal    & $10^9$          & 0          & The ``Toy Story'' effect \\
\ref{sec:terminal}     & Terminal epoch        & Temporal    & $\infty$        & $\infty$   & Survey ends first \\
\ref{sec:chipgap}      & Chip-gap              & Pipeline    & $10^9$          & 0          & Cozy in a chip gap \\
\ref{sec:compute}      & Compute-budget        & Pipeline    & $10^9$          & 0          & Bad Slurm etiquette (third party) \\
\ref{sec:classified}   & Government satellites & Pipeline    & \hl{[REDACTED]} & \hl{[REDACTED]} & \hl{[REDACTED]} \\
\hline
                       & Total                 &             & $\infty$        & 0--$\infty$ & \\
\hline
\end{tabular}
\vspace{0.5em}
\begin{minipage}{\linewidth}
\small
$^a$ Number of simulated objects \\
$^b$ Number of non-detected objects \\[0.3em]
\textit{Note.} We don't know how many objects are on fire in the solar system because no one will give us JWST time to look. We also don't know how many zero-velocity objects there are to model, but we also don't care because it's Gaia's problem not ours. The terminal-epoch population is formally infinite and is excluded from the total. The classified satellite column entries are classified.
\end{minipage}
\end{table*}

\begin{acknowledgments}
We thank the Sun for its continued operation throughout the survey period. Any interruption would affect these figures significantly and would also affect everything else.
\end{acknowledgments}

\begin{contribution}
All authors contributed equally to this, but some more equally than others.
\end{contribution}

\bibliography{sample701}{}
\bibliographystyle{aasjournalv7}

\end{document}